\documentclass[fleqn,usenatbib]{mnras}

\usepackage{newtxtext,newtxmath}
\usepackage[T1]{fontenc}
\usepackage{ae,aecompl}

\usepackage{amsmath}
\usepackage{amssymb}
\usepackage{epic}
\usepackage{graphicx}
\usepackage{hyperref}
\usepackage{multirow}
\usepackage{natbib}
\usepackage{overpic}
\usepackage{color}
\AtBeginShipout{%
 \ifnum\value{page}>1 %
 \typeout{* Additional boxing of page `\thepage'}%
 \setbox\AtBeginShipoutBox=\hbox{\copy\AtBeginShipoutBox}%
 \fi
}

\title[Cosmology with powerful radio-loud AGNs]{Cosmology with powerful radio-loud AGNs}

\author[R. J. Turner and S. S. Shabala]{
Ross J. Turner$^{1}$\thanks{Email: turner.rj@icloud.com} and Stanislav S. Shabala$^{1}$\\
$^{1}$School of Physical Sciences, University of Tasmania, Private Bag 37, Hobart, 7001, Australia}

\date{Accepted 2019 March 27. Received 2019 March 27; in original form 2018 November 2.}

\pubyear{2018}

\begin{document}

\label{firstpage}
\pagerange{\pageref{firstpage}--\pageref{lastpage}}
\maketitle

\begin{abstract}

Immensely bright quasars and radio-loud active galactic nuclei (AGNs) provide an enticing opportunity to construct standard candles detectable up to the very early universe. An analytic theory is proposed to measure the distance to powerful \citeauthor{FR+1974} type-II radio sources based on their integrated flux density across a broad range of radio frequencies, and the angular size and axis ratio of their synchrotron-emitting lobes. This technique can be used at low-redshift to construct absolute standard candles in conjunction with X-ray observations of the host cluster, or at high-redshift to measure the relative distances of objects and constrain the curvature of our universe. Distances calculated with this method are consistent for dissimilar objects at the same redshift; the two lobes of Cygnus A have flux densities, linear sizes and spectral break frequencies varying by between 15-35\% yet their fitted distances are the same to within 7\%. These distance estimates together yield a transverse comoving distance to Cygnus A of $261_{-55}^{+70}\rm\, Mpc$ corresponding to a Hubble constant of $H_0 =  64_{-13}^{+17}\rm\, km\, s^{-1}\, Mpc^{-1}$. Large samples of suitable FR-II sources could provide a measure of the Hubble constant independent of existing techniques such as the cosmic microwave background, baryon acoustic oscillations, and type 1a supernovae.

\end{abstract}

\begin{keywords}
cosmology: cosmological parameters -- cosmology: distance scale -- galaxies: active -- galaxies: jets -- radio continuum: galaxies
\end{keywords}

\section{INTRODUCTION}
\label{sec:INTRODUCTION}

Following the discovery of the expanding universe by \citet{Hubble+1929} in the early twentieth century, cosmologists concentrated on measuring a slowing of the so-called Hubble expansion expected from the gravitational attraction of matter. However, in 1998 two teams studying distant type 1a supernovae, the \emph{Supernova Cosmology Project} and the \emph{High-z SN Search}, independently concluded that expansion has been speeding up for the past $5\rm\, Gyr$ \citep{Riess+1998, Perlmutter+1999}. The strong observational evidence for cosmic acceleration, and thus indirectly Einstein's constant $\Lambda$, has led to the adoption of the $\Lambda$CDM model \citep[Lambda cold dark matter; e.g.][]{Spergel+2007, Planck+2016}. Evidence for an accelerating universe, and constraints on the energy and matter content of our universe, have only strengthened since the original supernova measurements. Other observational techniques including anisotropies in the cosmic microwave background (CMB), baryon acoustic oscillations, optical quasars, and the gas fraction in clusters, together place very tight, and largely consistent, constraints on our cosmology.

The Hubble constant is still most readily constrained through direct distance measurements to nearby objects, calibrated based upon known distances to closer objects (the cosmic distance ladder). \citet{Riess+2011} used \emph{Hubble Space Telescope} (HST) observations of type 1a supernovae with Cepheid variable distance anchors to measure the Hubble constant as $H_0 = 73.8\pm2.4 \rm\, km\,s^{-1}\, Mpc^{-1}$. These results have since been reanalysed by \citet{Efstathiou+2014} with a revised distance to the maser-host galaxy NGC\,4258 leading to an estimate of $70.6\pm3.3 \rm\, km\,s^{-1}\, Mpc^{-1}$. \citet{Riess+2011} also use Milky Way and Large Magellanic Cloud Cepheids as alternative distance anchors to NGC\,4258 giving a higher $73.9\pm2.7 \rm\, km\,s^{-1}\, Mpc^{-1}$. However, these Cepheid and type 1a supernovae based estimates are in tension with the CMB measurements: the \emph{Planck} CMB lensing and temperature data \citep{Planck+2016} finds a Hubble constant of $67.8\pm0.9 \rm\, km\,s^{-1}\, Mpc^{-1}$, statistically inconsistent with the supernovae results. Combining these measurements with the BOSS baryon acoustic oscillations observations from galaxy clustering and the $z \sim 3$ Lyman-$\alpha$ forest gives an even lower value of $66.4_{-0.9}^{+1.5} \rm\, km\,s^{-1}\, Mpc^{-1}$ \citep{Aubourg+2015}. Independent techniques not based on either a cosmic distance ladder or primordial anisotropies in the density and temperature may help resolve the reasons behind these differences.

Quasars and radio-loud active galactic nuclei (AGNs) have provided a tantalising opportunity to construct standard candles for use in cosmology since their discovery over five decades ago. In particular, radio AGNs are very numerous with up to 100 per cent of all large galaxies hosting an active nucleus at their centre \citep{Sabater+2018}, whilst their high luminosities can be detected from the present epoch out to the early universe at $z > 7$. These characteristics make AGNs ideal tools for probing our cosmology if their flux density, size or emission lines are standardisable. \citet{Watson+2011} showed the distance to $z < 0.3$ quasars can be measured using the known relation between the optical luminosity and the size of the broad emission line region \citep[see also][]{Haas+2011, Czerny+2013, King+2014}. The time lag between the optical and dust continuum has also been found to correlate with luminosity providing another technique to standardise optical quasars \citep[e.g.][]{Oknyanskij+1999, Oknyanskij+2001, Honig+2014, Yoshii+2014}, though the time lags can only reach redshifts up to $z \sim 1$ with current generation telescopes \citep[e.g.][]{Honig+2017}. 

There is a long history of techniques in the literature which attempt to standardise radio-loud AGNs. The \emph{Extended Radio Galaxy} method of \citet{Daly+1994} constructs standard rulers from the linear sizes of powerful radio sources on the assumption that the available energy to power their jets is independent of redshift; this argument is based on the observation that their median length is roughly constant out to $z = 0.5$ \citep[see also][]{Daly+2009}. This assumption ignores the changing FR-II environments with redshift (protoclusters versus poor groups), and the known difficulty in detecting old sources at high-redshift due to strong inverse-Compton losses from the CMB radiation \citep{BR+1999}. Meanwhile, \citet{Kellermann+1993} used VLBI observations to show the angular size--redshift relationship for a sample of 79 ultra compact radio sources was compatible with the deceleration parameter in a Friedmann cosmology. However, a correlation between the linear size and radio luminosity introduces a bias towards smaller objects at high redshifts; attempts were made to correct for this effect for both ultra compact \citep{Jackson+2004} and extended sources \citep{Buchalter+1998}. As a result, radio AGN standard candles are presently not in wide acceptance, which is particularly unfortunate in light of the numerous high-sensitivity large-sky radio-frequency surveys currently in progress (e.g. ASKAP EMU, \citealt{Norris+2011}; ASKAP POSSUM, \citealt{Gaensler+2010}; LOFAR LoTSS, \citealt{Shimwell+2017, Shimwell+2019}; MWA GLEAM, \citealt{Wayth+2015}).

The analytic theory underpinning the \emph{Radio AGNs in Semi-analytic Environments} (RAiSE; \citealt{Turner+2015}; \citealt{Turner+2018a}) model for the dynamical and synchrotron evolution of active radio galaxies is modified in this work to physically link distinct observations of radio lobes with each other (i.e. multi-frequency flux density and angular size). The existing form of the model has been shown to: reproduce the surface brightness and spectral age maps of 3C31 and 3C436 \citep{Turner+2018a}; measure jet kinetic powers consistent with X-ray inverse-Compton measurements \citep{Turner+2018b}; infer duty-cycles in episodic sources comparable to the observed radio-loud fraction \citep{Turner+2018}; and show dynamical evolution in the lobe length, axis ratio and volume consistent with hydrodynamical simulations \citep{Turner+2018b}. The standardising of radio-frequency AGNs may provide an independent and readily detectable distance measure across a broad redshift range (i.e. $0 < z < 7$) to constrain the Hubble constant (in addition to the matter and energy densities), and to resolve the tension between the CMB and supernovae measurements.

In this paper, we derive an equation for the distance to radio-loud AGNs enabling observations of these objects to be standardised and hence the construction of standard candles (Section \ref{sec:Constructing radio AGN standard candles}). The distance equation is modified in Section \ref{sec:Measuring the host cluster environment} to provide absolute distance measurements by calibrating the terms describing the properties of the radio source host environment. This technique is used to measure the distance to Cygnus A and thus derive the Hubble constant (Section \ref{sec:Distance to Cygnus A}), then in Section \ref{sec:High-redshift radio source catalogue} we briefly describe the required properties of any additional FR-II radio galaxies to which this method can be applied.

The spectral index $\alpha$ is defined throughout the paper in the form $S = \nu^{-\alpha}$ for flux density $S$ and frequency $\nu$.

\section{Constructing radio AGN standard candles}
\label{sec:Constructing radio AGN standard candles}

The giant radio lobes emanating from the nuclei of active galaxies are well modelled, both analytically \citep[e.g.][]{KA+1997, Blundell+1999, Luo+2010, Turner+2015, Hardcastle+2018} and numerically \citep[e.g.][]{Hardcastle+2014, English+2016, Yates+2018, Massaglia+2019}, enabling intrinsic parameters such as the source age and kinetic power to be estimated from observables. In this section, we develop a technique for measuring the distance to powerful \citeauthor{FR+1974} type-II (FR-II) radio sources based primarily on radio observables including their flux density, angular size and spectral energy distribution.

\subsection{Distances in the $\Lambda$CDM cosmology}

The standard concordance model of Big Bang cosmology is the $\Lambda$CDM model, in which the universe contains dark energy and matter at densities $\Omega_{\rm \Lambda}$ and $\Omega_{\rm m}$ respectively. Distance measures depend crucially on the values of the cosmological parameters; the Hubble constant $H_0$ describes the expansion rate of the universe whilst the energy and matter densities parametrise its shape.

The Hubble distance is defined in terms of the Hubble constant $H_0$ and speed of light $c$ through the relation
$d_{\rm H} = c/H_0$. The comoving distance, the distance between two objects as measured at the present cosmological time, is related to the Hubble distance $d_{\rm H}$ through

\begin{equation}
d_{\rm C}(z) = d_{\rm H}\int_0^z \frac{dz'}{\sqrt{\Omega_{\rm m}(1 + z')^3 + \Omega_{\rm k}(1 + z')^2 + \Omega_\Lambda}} ,
\end{equation}

where the denominator is the cosmic evolution of the Hubble constant (i.e. $E(z) = H(z)/H_0$), $\Omega_{\rm k} = 1 - \Omega_{\rm m} - \Omega_{\rm \Lambda}$ describes the curvature of the universe, and $z$ is the redshift. In a flat universe $\Omega_{\rm k} = 0$, for an open (hyperbolic) universe $\Omega_{\rm k} > 0$ and for a closed (spherical) universe $\Omega_{\rm k} < 0$. This curvature of the universe is considered in the modified form of the comoving distance, the transverse comoving distance:

\begin{equation}
d_{\rm M}(z) = 
\begin{cases}
\frac{d_{\rm H}}{\sqrt{\Omega_{\rm k}}}\sinh(\sqrt{\Omega_{\rm k}}d_{\rm C}(z)/d_{\rm H}) &\text{if }\Omega_{\rm k} > 0\\
d_{\rm C}(z) &\text{if }\Omega_{\rm k} = 0\\
\frac{d_{\rm H}}{\sqrt{\Omega_{\rm k}}}\sin(\sqrt{-\Omega_{\rm k}}d_{\rm C}(z)/d_{\rm H}) &\text{if }\Omega_{\rm k} < 0 .
\end{cases} 
\label{comoving}
\end{equation}

The angular diameter and luminosity distances are directly related to this distance and the redshift as

\begin{equation}
\begin{split}
d_{\rm A}(z) &= \frac{d_{\rm M}(z)}{1 + z}\\
d_{\rm L}(z) &= (1 + z) d_{\rm M}(z) .
\end{split}
\end{equation}

These two distances may be constrained for an observed radio source if its physical size $D$ and luminosity $L_\nu$ are known, where $\nu$ is the observing frequency. The measured angular size $\theta$ and flux density $S_\nu$ of a lobe are related to their intrinsic counterparts through

\begin{equation}
\begin{split}
D &= {d_{\rm A}}(z) \theta = \frac{d_{\rm M}(z) \theta}{1 + z}\\
L_\nu &= 4\pi {d_{\rm L}}^2(z) S_\nu = 4\pi (1 + z)^2 {d_{\rm M}}^2(z) S_\nu .
\end{split}
\label{trans}
\end{equation}

The physical lobe size and luminosity map to the redshift and transverse comoving distance with different functional dependencies; hence their combination can be used to constrain either redshift or the cosmology. The complication is that the relationship between size and luminosity varies as the lobes evolve throughout the evolutionary history of the radio source.

\subsection{Luminosity--size relationship from analytic radio galaxy models}
\label{sec:Luminosity--size relationship from analytic AGN models}

\begin{figure*}
\begin{center}
\includegraphics[width=0.46\textwidth,trim={5 5 5 5},clip]{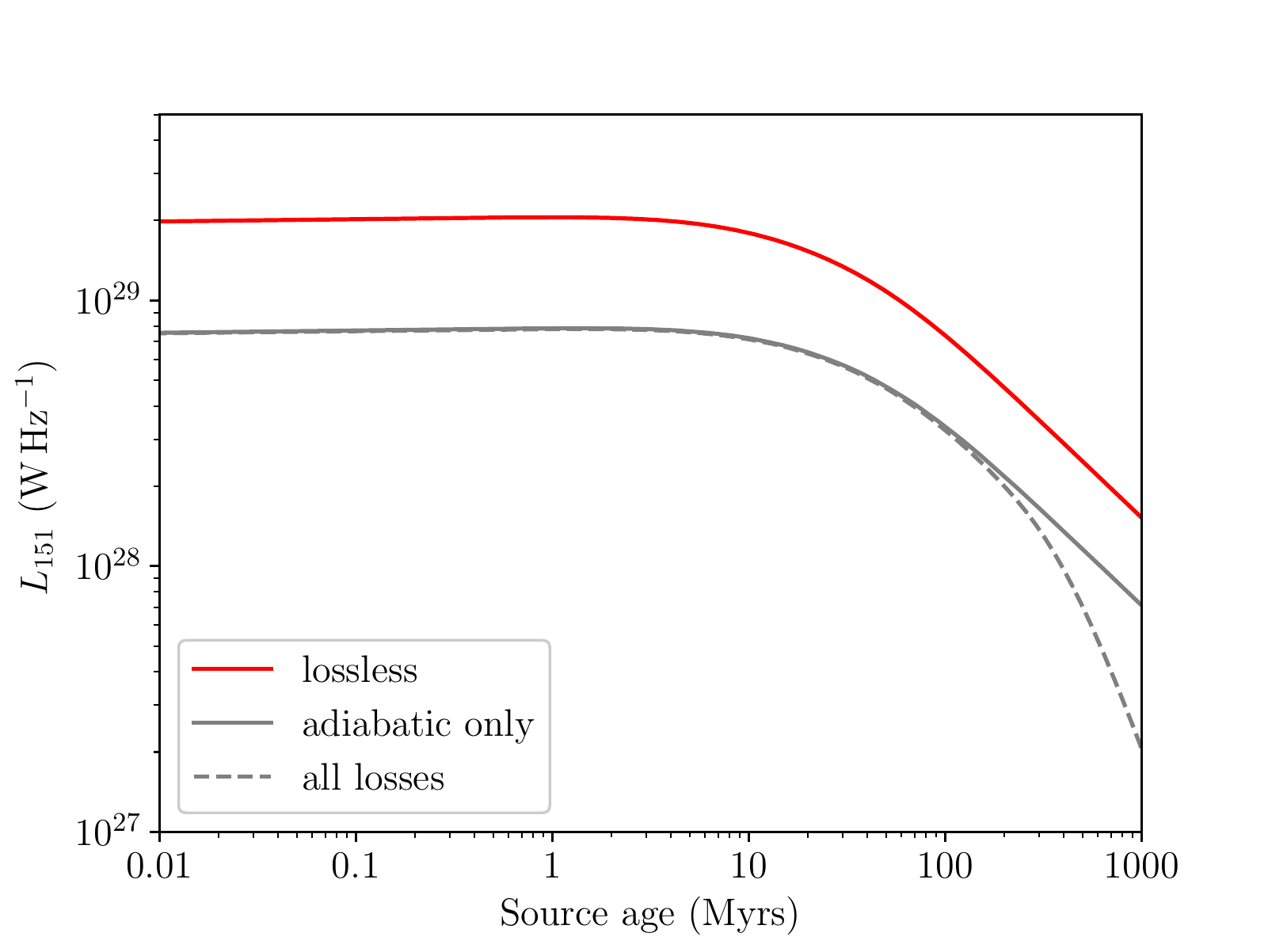}\includegraphics[width=0.46\textwidth,trim={5 5 5 5},clip]{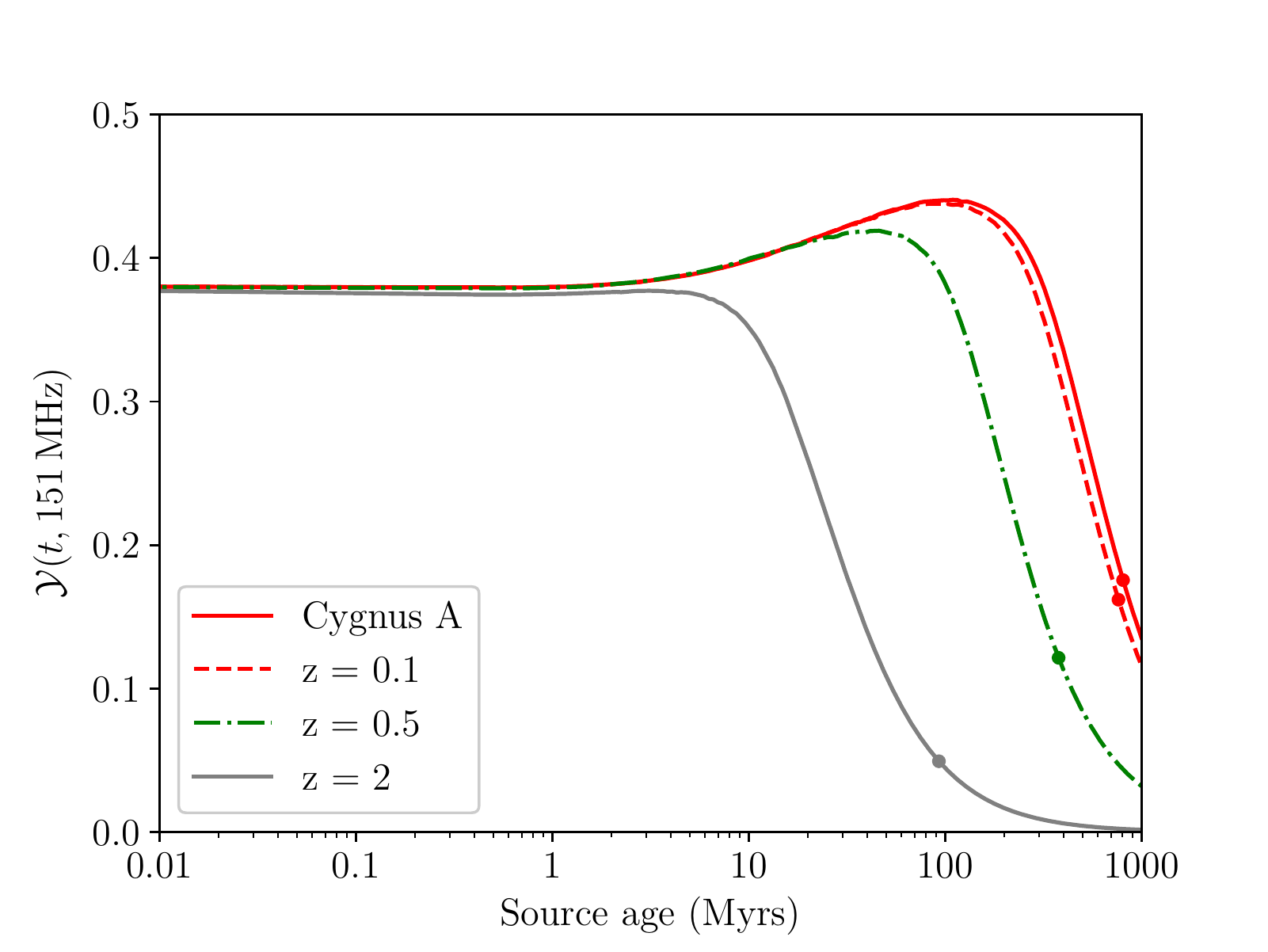}
\end{center}
\caption{Radio source luminosity at $151\rm\, MHz$ for both lobes of a Cygnus A-like source as a function of the age of the source (left panel). The simulated radio source \citep[using RAiSE;][]{Turner+2015} has the same luminosity, linear size, axis ratio, gas density profile and properties of the electron population (i.e. $q$, $s$ and $\gamma$) as observed in Cygnus A. The lossless luminosity is shown in red, a model including losses from the adiabatic expansion of the lobe is shown in solid grey, and the luminosity including all {loss mechanisms} is shown in dashed grey. The synchrotron and inverse-Compton losses only begin to affect the level of emission at this frequency beyond $100\rm\, Myr$. The loss function $\mathcal{Y}(t, \nu)$ is plotted in the right panel for Cygnus A (at $z = 0.056$) and Cygnus A-like sources at redshift $z = 0.1$, 0.5 and 2. The inverse-Compton losses are more severe at $z = 2$ and affect the level of emission at $151\rm\, MHz$ from $10\rm\, Myr$. The maximum age the Cygnus A-like sources can be detected in LoTSS \citep[for surface brightness above 5$\sigma$ level;][]{Shimwell+2019} are shown by dots towards the right-hand end of the tracks.}
\label{fig:LDtracks}
\end{figure*}

The injection of particles into the radio lobe produces an initial power law distribution of electron energies $N(E) = N_0 E^{-s}$, where $N_0$ is a constant and $s = 2\alpha_{\rm inj} + 1$ for injection-time spectral index $\alpha_{\rm inj} > 0.5$. The population of synchrotron-emitting electrons has a minimum and maximum particle energy $\gamma_{\rm min}$ ($\equiv\!\!\gamma$) and $\gamma_{\rm max}$ respectively. The luminosity arising from the synchrotron-emitting electrons in the lobes of powerful radio sources is a function of the observing frequency $\nu$, energetics of the electron population, and the shape and pressure of the lobe. {From Equations 3, 12 and 13 of \citet{Alexander+2000}, we may derive an expression} for the luminosity of self-similar FR-II radio lobes expanding into a power law environment (i.e. $\rho \propto r^{-\beta}$):

\begin{equation}
\begin{split}
L_\nu &= \frac{4\pi f_1(s) \nu^{-(s - 1)/2}}{A^2 \gamma^{2 - s}}\left(\frac{f_2(\beta) a^\beta}{\Gamma_{\rm c} - 1} \frac{q \rho}{q + 1}\right)^{(s + 5)/4} t^{-(s + 5)/2}\\
&\quad\times\, D^{(22 + 2s - s\beta - 5\beta)/4} ,
\label{lumin1}
\end{split}
\end{equation}

where $D$ is the lobe linear size, $t$ is the source dynamical age, $\rho$ is the density of the host galaxy environment at an arbitrary scale radius $a$ (set equal to the lobe length), and $q$ is the ratio of energy in the magnetic field to that in the particles. Here, $A$ is the axis ratio of the lobe (i.e. radio lobe length divided by its cross-sectional radius). 
The constants of proportionality $f_1$ and $f_2$ are functions of $s$ and $\beta$ respectively defined as

\begin{equation}
\begin{split}
&f_1(s) = \frac{\sigma_{\rm T} (s - 2)}{9 m_{\rm e} c}\left(\frac{e^2 \mu_0}{2 \pi^2 {m_{\rm e}}^2}\right)^{(s - 3)/4} \mathcal{Y}(t, \nu) \\
&f_2(\beta) = \frac{18 \chi}{(\Gamma_{\rm x} + 1)(5 - \beta)^2} ,
\end{split}
\label{kappa1}
\end{equation}

where $\sigma_{\rm T}$ is the electron scattering cross-section, $e$ and $m_{\rm e}$ are the electron charge and mass, $c$ is the speed of light, $\mu_0$ is the vacuum permeability, $\Gamma_{\rm x} = 5/3$ is the adiabatic index of the external medium, and $\chi$ is the ratio of the lobe to expansion surface pressures which is typically of order unity \citep[e.g. $6\times10^{-12}\rm\, Pa$ and $10^{-11}\rm\, Pa$ respectively for Cygnus A;][]{Carilli+1996}. Finally, $\mathcal{Y}(t, \nu)$ is the loss function defined in Equation 4 of \citet{Turner+2018a}. The synchrotron emissivity is simulated using RAiSE for a radio source with properties of the lobes and cluster environment in broad agreement with observations of Cygnus A (see Figure \ref{fig:LDtracks}). When observed at frequencies below the optically-thin spectral break the {loss function is} dominated by adiabatic expansion leading to an approximately constant value in the range $\mathcal{Y} = 0.3$-0.5 for the Cygnus A-like source. The synchrotron and inverse-Compton losses at $151\rm\, MHz$ {are less significant than the adiabatic losses} for radio sources younger than $100\rm\, Myr$ except at high-redshift ($z \gg 0.5$) where the stronger microwave background radiation leads to {significant} losses in $10\rm\, Myr$ old sources.

The lobe pressure of the FR-II is similarly derived in terms of the source age, size and host environment as (c.f. dynamical model in \citealt{Turner+2015})

\begin{equation}
p = f_2(\beta) \rho a^\beta t^{-2} D^{2 - \beta} ,
\label{press1}
\end{equation}

The lobe magnetic field strength $B = \sqrt{2\mu_0 u_{\rm B}}$ follows directly from this equation and the relationship between the pressure and energy density of the magnetic field $u_{\rm B}$ \citep[Equation 15 of][]{KDA+1997}. That is,

\begin{equation}
\begin{split}
B = \left(\frac{2\mu_0}{\Gamma_{\rm c} - 1} f_2(\beta) a^\beta \frac{q \rho}{q + 1}\right)^{1/2} t^{-1} D^{(2 - \beta)/2} ,
\label{field1}
\end{split}
\end{equation}

where $\Gamma_{\rm c} = 4/3$ is the adiabatic index of the radio lobe plasma. The relationship between the physical size and luminosity as it stands is a function of variables which can largely be constrained observationally, or statistically using a prior probability distribution. The exception to this is the source dynamical age, which cannot be directly measured and can vary by several orders of magnitude in observed sources.
\citet{Turner+2018a, Turner+2018b} use hydrodynamical simulations tracing the synchrotron-emitting electron population to show that the spectral age fitted by continuous injection models is equal to the dynamical age. The discrepancy observed between dynamical and Jaffe-{Perola} \citep[JP;][]{Jaffe+1973} model spectral ages is explained by the mixing of different age electrons throughout lobed sources, thus violating the impulsive injection assumption in the JP model \citep{Turner+2018a}, or through inhomogeneous magnetic fields \citep{Hardcastle+2013}. The spectral age, and thus dynamical age, of the radio source is related to the break frequency in the observer frame $\nu_{\rm b}$ through

\begin{equation}
\tau = \frac{\upsilon B^{1/2}}{\langle B^2 \rangle + {B_{\rm ic}}^2}\left[\nu_{\rm b} (1 + z) \right]^{-1/2} = \frac{\upsilon B^\sigma}{\kappa^2} \left[\nu_{\rm b} (1 + z) \right]^{-1/2},
\label{spectral}
\end{equation}

where $B$ is the lobe magnetic field strength, $B_{\rm ic} = 0.318 (1 + z)^2 \rm\, nT$ is the magnitude of the magnetic field equivalent to the microwave background, and the constant $\upsilon$ is defined in Equation 5 of \citet{Turner+2018b}. {The time-averaged field strength for the presently emitting electrons ${\langle B\rangle}$ is expected to be at most 5-10\% higher than the instantaneous value (i.e. ${\langle B\rangle} \sim B$) based on RAiSE simulations which show the most recently injected 10-15\% of electrons contribute 50\% of the flux density at standard observing frequencies.} The dependence on the magnetic field strength in the fraction on the left-hand side of the equality can be expressed locally as a power law to enable an analytic solution to be maintained. The constants $\sigma$ and $\kappa$ can be derived from Equation \ref{spectral} yielding the following expressions:

\begin{subequations}
\begin{equation}
\sigma = \frac{-3 B^2 + {B_{\rm ic}}^2}{2(B^2 + {B_{\rm ic}}^2)} ,
\label{sigma}
\end{equation}
\begin{equation}
\kappa = \left[B^{\sigma - 1/2} (B^2 + {B_{\rm ic}}^2) \right]^{1/2} .
\end{equation}
\end{subequations}

In the limit of weak lobe magnetic fields ($B \ll B_{\rm ic}$) these constants have values of $\sigma = 1/2$ and $\kappa = B_{\rm ic}$, whilst in the strong field limit ($B \gg B_{\rm ic}$) they have values of $\sigma = -3/2$ and $\kappa = 1$. 

Substituting the lobe field strength of Equation \ref{field1} into Equation \ref{spectral} gives an expression for the source dynamical age as a function of size and known parameters:

\begin{equation}
\begin{split}
t = \left[\left(\frac{2\mu_0}{\Gamma_{\rm c} - 1} f_2(\beta) a^\beta \frac{q \rho}{q + 1} D^{2 - \beta}\right)^{-\sigma} \frac{\nu_{\rm b} (1 + z) \kappa^4}{\upsilon^2}\right]^{-1/2(\sigma + 1)} .
\label{age1}
\end{split}
\end{equation}

The radio lobe luminosity and physical size can thus be related to each other and the spectral break frequency observable by substituting this Equation \ref{age1} into Equation \ref{lumin1}. That is,

\begin{equation}
\begin{split}
L_\nu = \frac{4\pi f_1(s)\, \nu^{-(s - 1)/2}}{A^2 \gamma^{2 - s}}\left(f_3(\beta) a^\beta \frac{q \rho}{q + 1}\right)^x [\nu_{\rm b} (1 + z)]^x D^y ,
\label{lumin2}
\end{split}
\end{equation}

where we have collected the constants in the magnetic field and break frequency terms together as

\begin{equation}
f_3(\beta) = \frac{f_2(\beta) \kappa^4}{(2\mu_0)^{\sigma}(\Gamma_{\rm c} - 1)\upsilon^2} .
\end{equation}

The exponents $x$ and $y$ are constant functions of $s$ and $\beta$ defined as $x = (s + 5)/4(\sigma + 1)$
and $y = (22 + 12\sigma + 2s - 5\beta - s\beta)/4(\sigma + 1)$. Here, the exponent of the electron energy distribution takes values between $s = 2$ and 3, whilst the slope of the host gas density profile steepens from $\beta \gtrsim 0$ near the core to $\beta = 2$ at larger radii. 

The expressions in Equation \ref{trans} for the radio lobe luminosity and physical size can now be coupled using Equation \ref{lumin2}. We thus arrive at a relation for the transverse comoving distance in terms of the lobe flux density, angular size, spectral break frequency and redshift observables. That is,

\begin{equation}
\begin{split}
d_{\rm M}(z) &= \bigg[\frac{A^2 \gamma^{2 - s}\nu^{(s - 1)/2} S_\nu}{f_1(s)}\bigg]^{-1/(2 - y)}\left(f_3(\beta) a^\beta \frac{q \rho}{q + 1}\right)^{x/(2 - y)}\\
&\quad\times\, {\nu_{\rm b}}^{x/(2 - y)}\theta^{y/(2 - y)} (1 + z)^{(x - y - 2)/(2 - y)} .
\label{dM}
\end{split}
\end{equation}

For typical values of $s$ and $\beta$, the transverse comoving distance is strongly related to the angular size $\theta$, break frequency $\nu_{\rm b}$, spectroscopic redshift $z$, gas density $\rho$ and equipartition factor $q$. By contrast, it has a weaker dependence on the flux density $S_\nu$, shape of the lobe $A$, and properties of the electron distribution (i.e. $\gamma$). These dependences are investigated fully in Section \ref{sec:Measuring the host cluster environment} for a modified form of this equation.

\subsection{Applications of analytic AGN standard candles}
\label{sec:Applications of analytic AGN standard candles}

The theory presented in this section to measure the transverse comoving distance to radio-loud AGNs needs to be modified for particular applications. In particular, the gas density in the host atmosphere and magnetic field strength in the lobe plasma are not readily observable quantities, especially at high-redshift. Potential applications of these AGN standard candles and the necessary modifications to the distance equation include:

\begin{itemize}
	\item Absolute standard candles: the density, equipartition factor and Lorentz factor of the electron population are calibrated for each source individually or using a population average based on X-ray and radio frequency observations. X-ray observations of cluster environments are rare beyond $z \sim 0.5$ and are generally for a targeted sample, not as part of large sky surveys \citep[e.g.][]{Croston+2005, Ineson+2017}. This technique will likely only constrain the Hubble constant.
	\item Relative distance measurements: the uncertain intrinsic properties of AGNs are assumed to come from a random normal distribution, which is independent of redshift. The cosmic evolution of galaxies may be modelled over a broad redshift range using semi-analytic galaxy evolution models \citep[e.g. SAGE;][]{Croton+2016} for a trial cosmology to minimise systematic uncertainties in the gas density. This technique is capable of measuring the curvature in the redshift--distance relationship at high-redshift (i.e. $z \gg 1$) to constrain the matter and dark energy densities.
	\item Radio continuum redshifts: the distance equation can be calibrated using radio AGNs with spectroscopic redshifts (and assuming a cosmology) to constrain the distance to sources lacking redshifts.
Radio polarimetry may be used to measure the Faraday rotation of polarised emission from the lobes to constrain the relative quantities of gas surrounding different radio galaxies; this technique is likely only viable up to moderate redshifts (i.e. $z\sim 0.3$) to ensure multiple beams across each source \citep[e.g. $10\rm\, arcsec$ for POSSUM;][]{Gaensler+2010}. The distribution of X-ray observed gas densities in well-studied galaxies can alternatively be used to statistically determine the most probable range of redshifts using Bayesian statistics.
\end{itemize}

In this work, we proceed to investigate how the analytic AGN distance measure of Equation \ref{dM} is transformed into an absolute distance measurement using X-ray and radio frequency observations of the gas density in clusters, typical equipartition factors, and the minimum Lorentz factor of the electron population. The ability of high-redshift AGNs to constrain the matter and dark energy densities using the relative distance measurement technique with the full RAiSE model (to increase precision in the modelling) will be explored in a subsequent publication.

\section{Measuring the AGN host environment}
\label{sec:Measuring the host cluster environment}

The AGN standard candle distance equation proposed in the previous section includes the gas density of the cluster environment as a key parameter; this must either be constrained observationally to provide absolute distance measurements, or at least the relative densities of different clusters characterised if using the standard candles to derive redshifts or the dark energy and matter densities. The distance equation is modified in this section assuming X-ray observations of the gas density will be used to provide absolute distance measurements.

\subsection{X-ray density in clusters}
\label{sec:X-ray density in clusters}

The gas density in clusters is generally measured using radial profiles of the projected temperature and X-ray surface brightness \citep[see e.g.][]{Vikhlinin+2005}. The quantised nature of the X-ray surface brightness (photon counts per solid angle) means observations are often smoothed, either by applying a smoothing function over the measurements or by summing the counts in annuli about the cluster centre. The fraction of the telescope effective area covered by each pixel or annulus is calculated enabling the counts originating from that region to be determined. The gas temperature and metallicity are fitted from the X-ray spectrum in each pixel or annulus, e.g. using the {\sc mekal} model for hot, diffuse gas emission \citep{Mewe+1985, Mewe+1986, Kaastra+1992, Liedahl+1995}.  Importantly, the effective area, and the projected temperature and metallicity are invariant to the assumed distance to the radio galaxy. These three observables are used to convert the X-ray count rate into the emission measure, $E\!\:\,\!\!M$, along the lines-of-sight passing through the pixel or annulus. That is, the normalisation to the X-ray spectra for a pixel or annulus at cylindrical radius $\theta$ from the cluster centre (in the plane of the sky) is fitted by the emission measure of the form

\begin{equation}
\begin{split}
E\!\:\,\!\!M(\theta) = \frac{k_1 d\Omega \:\! d_{\rm M}(z)}{4\pi (1 + z)^3} \int n^2(\theta, l) dl ,
\end{split}
\label{emission_integral}
\end{equation}

where $l$ is the angular distance along the line-of-sight, $a = \sqrt{\theta^2 + l^2}$ is the spherical radius from the galactic centre, $d\Omega$ is the solid angle of the pixel or annulus, and $k_1$ is a model dependent constant (e.g. $k_1 = 10^{-14} \rm\, cm^{6}\, pc^{-1}\, m^{-5}$ for {\sc mekal}). The number density of ions is defined through $n^2(\theta, l) = n_{\rm p} n_{\rm e}$ where $n_{\rm p}$ and $n_{\rm e}$ are the number densities of positive and negative ions respectively.

The gas density at an angular distance $\theta$ from the cluster centre is found from the number density as $\rho(\theta) = 0.6 m_{\rm p} n(\theta, 0)$ for an average ion mass $0.6 m_{\rm p}$. The density at an arbitrary radius $a$ (which we set to be the scale radius defined earlier) is derived by extrapolating the density profile ($\rho \propto r^{-\beta}$) from the measured value at an angular radius $\theta$. That is,

\begin{equation}
\begin{split}
\rho(a) = \eta \left( \frac{a}{\theta} \right)^{-\beta} {d_{\rm M}}^{\beta - 1/2}(z) (1 + z)^{3/2 - \beta} ,
\end{split}
\label{kband_density2}
\end{equation}

where $\eta \equiv \eta(\theta)$ is a redshift and cosmology independent form of the gas density at the end of the lobe, $\theta$ is defined in radians and the scale radius $a$ has the same units as the transverse comoving distance. The de-distanced gas density (i.e. $\eta$) is largely independent of the exponent of the gas density profile (assuming for any trial cosmology that $a$ still approximately corresponds to $\theta$) with only a weak dependence on redshift of the form $(1 + z)^{-\beta}$. This is inconsequential at low redshifts for typical varations in $\beta$ in the range $0.82\pm0.15$ \citep{Vikhlinin+2006}; e.g. for Cygnus A at $z = 0.056$ this leads to a less than 1\% uncertainty, though increases up to 5\% at $z=1$. {Further, the gas density at the end of the lobe may be overestimated by 3\% for a typical source inclined 30 degrees to the plane of the sky, and up to 12\% for an angle of 45 degrees (since $\rho \propto r^{-\beta}$, also see discussion in Section \ref{sec:Stability of method with magnetic field strength}).}

\subsection{Absolute AGN standard candles}

The gas density has been recast in terms of directly observable quantities (i.e. the de-distanced X-ray gas density) enabling analytic AGN standard candles to be constructed from directly measurable properties of radio sources and their host galaxies. The transverse comoving distance in Equation \ref{dM} is reformulated using the relation for the de-distanced X-ray gas density in Equation \ref{kband_density2}. That is,

\begin{equation}
\begin{split}
d_{\rm M}(z) &= \bigg[\frac{A^2 \gamma^{2 - s}\nu^{(s - 1)/2} S_\nu}{f_1(s)}\bigg]^{-1/\delta} \left(f_3(\beta) \frac{q \eta}{q + 1} \right)^{x/\delta}\\
&\quad\times\, {\nu_{\rm b}}^{x/\delta} \theta^{(y + \beta x)/\delta} (1 + z)^{[(5/2 - \beta)x - y - 2]/\delta} ,
\label{dMenv}
\end{split}
\end{equation}

where the exponent $\delta = 2 - (\beta - 1/2)x - y = -(23 +8\sigma + 3s)/8(\sigma + 1)$ is a constant function of $s$. Importantly, the exponents on all terms in this equation have no dependence on the slope of the radio source environment $\beta$ upon simplifying each expression. That is, the distance equation has no dependence on the slope of the cluster density profile except in the denominator of the $f_3(\beta)$ term as $(5 - \beta)^2$. The distance measure is therefore expected to be robust to typical variations in this parameter \citep[e.g.][]{Vikhlinin+2006}. The stability of the distance measure to variations in the other parameters is investigated in the following section.

\subsection{Stability of method with magnetic field strength}
\label{sec:Stability of method with magnetic field strength}

The stability of the method is quantified by assessing the order of the exponents in Equation \ref{dMenv} for typical electron populations and lobe magnetic field strengths; i.e. $s$ and $\sigma$ respectively. Typical exponents are calculated for an injection electron energy distribution with slope at the 16 and 84th percentiles measured for FR-IIs in the 3CRR sample by \citet{Turner+2018b}, and the lobe magnetic field strength either equal to the field equivalent of the microwave background (i.e. $B = B_{\rm ic}$), or in the limit that lobe field strength is much greater than or much less than the CMB value (i.e. $B \gg B_{\rm ic}$ or $B \ll B_{\rm ic}$). The exponents on the key observables in Equation \ref{dMenv} are tabulated in Table \ref{tab:exponents} for each combination of these parameters.

\begin{table*}
\begin{center}
\caption[]{The variation in the exponents of the distance equation for magnetic field strengths equal to the field equavalent to the cosmic microwave background radiation and in the two limiting cases. The first column is the assumed magnetic field strength, the second is the corresponding value of $\sigma$, the third column is the assumed value for the electron energy injection index, and the fourth through eighth columns are the dependencies on the parameters in the distance equation.}
\label{tab:exponents}
\renewcommand{\arraystretch}{1.1}
\setlength{\tabcolsep}{8pt}
\begin{tabular}{cccccccccc}
\hline\hline
Magnetic field\!\!&\!\!\!\!\!&\multicolumn{2}{c}{Parameters}&\!\!\!\!\!&\multicolumn{5}{c}{Exponents}\\
	&\!\!\!\!\!&$\sigma$	&	$s$&\!\!\!\!\!&	$S_\nu$,\,$A^2$	&$\theta$	&$\nu_{\rm b}$,\,$\eta$,\,$q$	&$\gamma$	&$1+z$	
\\
\hline
\multirow{2}{*}{$B \ll B_{\rm ic}$}\!\!&\!\!\!\!\!&\multirow{2}{*}{0.5}	&	2.13&\!\!\!\!\!&	0.36&	-1.93&	-0.43&	-0.05&	1.58
\\
	&\!\!\!\!\!&	&	2.42&\!\!\!\!\!&	0.35&	-1.92&	-0.43&	-0.15&	1.53
\\
\multirow{2}{*}{$B = B_{\rm ic}$}\!\!&\!\!\!\!\!&\multirow{2}{*}{-0.5}	&	2.13&\!\!\!\!\!&	0.16&	-1.60&	-0.56&	-0.02&	0.51
\\
	&\!\!\!\!\!&	&	2.42&\!\!\!\!\!&	0.15&	-1.59&	-0.57&	-0.06&	0.48
\\
\multirow{2}{*}{$B \gg B_{\rm ic}$}\!\!&\!\!\!\!\!&\multirow{2}{*}{-1.5}	&	2.13&\!\!\!\!\!&	-0.23&	-0.95&	-0.82&	0.03&	-1.56
\\
	&\!\!\!\!\!&	&	2.42&\!\!\!\!\!&	-0.22&	-0.97&	-0.81&	0.09&	-1.50
\\
\hline
\end{tabular}
\end{center}
\end{table*}

The exponents on the majority of the observable parameters have absolute values much less than unity; i.e. the propagated fractional uncertainty in the distance measurement will be significantly smaller than the fractional measurement uncertainty. Notable exceptions are the linear size of the lobes and the spectroscopic redshift of the source for which measurement uncertainties can be at worst squared when propagated through to the distance calculation. However, both the linear size of the lobe and the spectroscopic redshift typically have some of the lowest measurement uncertainties, although there can be systematic uncertainties in the linear size resulting from the viewing angle and determining the maximal extent of the lobe in poorly resolved sources. 

\citet{Turner+2018b} argue the length of a typical radio source {(with $A \sim 4$ and approximately spherical lobe ends)} viewed at 30 degrees from normal to the jet axis would be observed as 97\% of its intrinsic length, reducing to 87\% for a viewing angle of 45 degrees. These uncertainties propagate through to a 3-6\% error in the distance measurement for a viewing angle of 30 degrees and a 12-27\% error for the more extreme 45 degree viewing angle. {However, the error in the gas density introduced by the inclination angle (see Section \ref{sec:X-ray density in clusters}) opposes the errors in the source size, reducing the uncertainties in the distance to 1-5\% and 4-24\% respectively (lower bound is for the $B \gg B_{\rm ic}$ limit). Thin or sharply pointed lobes are expected to be more sensitive to projection effects at smaller angles.} We note that radio sources at extreme viewing angles can be excluded {(resolution permitting)} based on either the relativisitic beaming of emission in their two jets or the strength of their core emission; the inclination angle is known to be anti-correlated with the fractional flux density in the core \citep{Orr+1982}. {However, \citet{Carilli+1991} use this technique to constrain the inclination angle Cygnus A as greater than 27 degrees with VLBI images of the parsec-scale jets, inconsistent with observations of the large-scale morphology by \citet{Dreher+1981} and \citet{Hargrave+1974} which suggest Cygnus A lies close to the plane of the sky ($\lesssim 15$ degrees).}

\begin{figure}
\begin{center}
\includegraphics[width=0.46\textwidth,trim={5 5 5 5},clip]{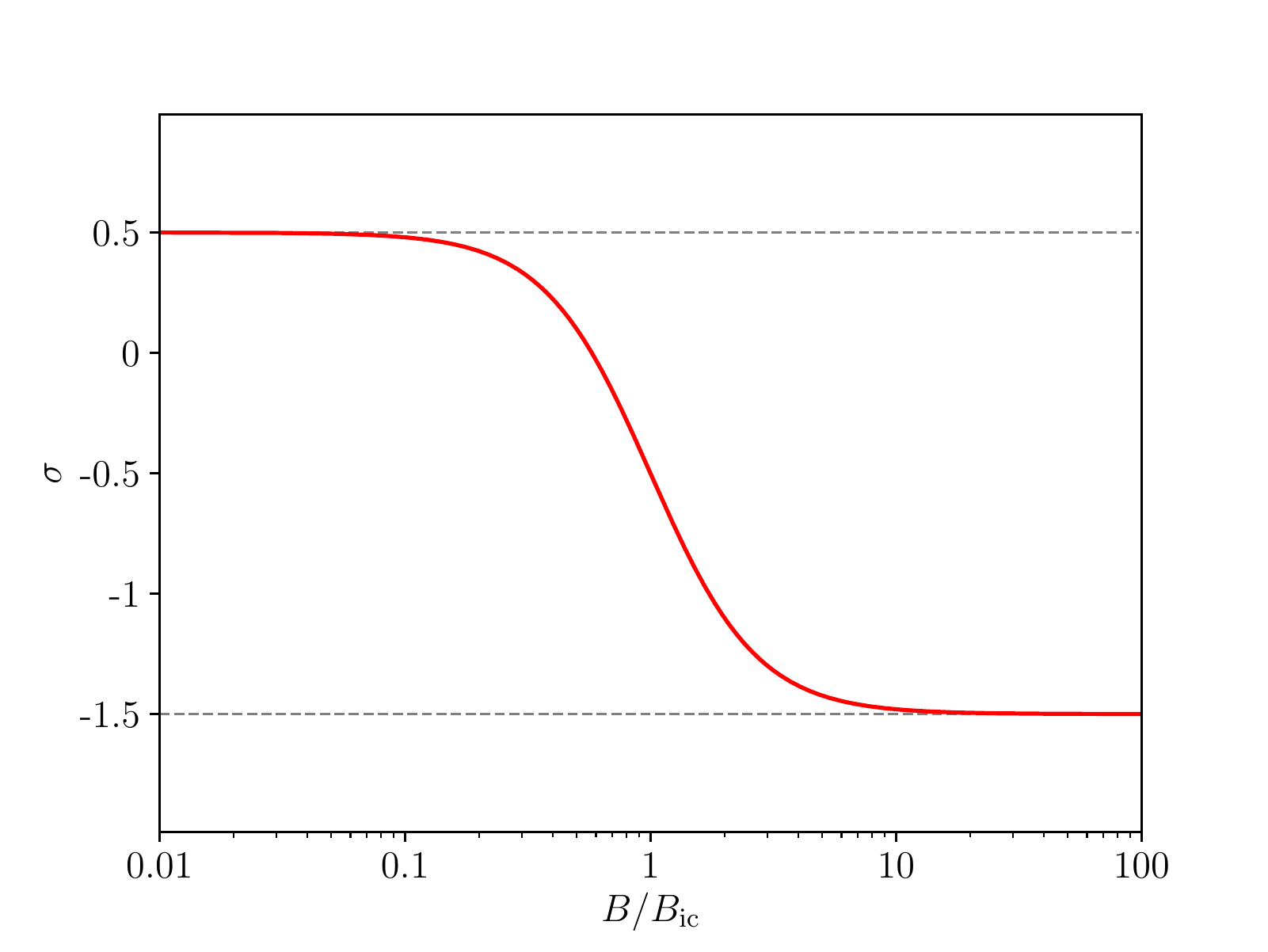}
\end{center}
\caption{The value of the exponent $\sigma$, defined in Equation \ref{spectral} to maintain an analytical solution, as a function of the ratio of the lobe magnetic field strength to the field equivalent of the cosmic microwave background radiation (solid red line). The dashed grey lines show the two limiting cases referred to as the strong and weak field limits.}
\label{fig:field_sigma}
\end{figure}

The exponents on all the parameters vary greatly as the lobe magnetic field changes from the limit of field strengths greatly below the field equivalent of the cosmic microwave background radiation to field strengths greatly above this value (i.e. from the $B \ll B_{\rm ic}$ limit to the $B \gg B_{\rm ic}$ limit). In particular, the exponents for the flux density and spectroscopic redshift change sign, whilst the linear size exponent changes from a linear to a quadratic dependence. The distance measure is therefore expected to be highly unstable if the value of $\sigma$ changes rapidly near the lobe magnetic field strength. From Equation \ref{sigma} it can be shown that $\sigma$ converges to its limiting values for magnetic field strengths outside approximately a factor of ten of the field equivalent of the cosmic microwave background radiation; i.e. $\sigma \approx 0.5$ if $B < 0.1B_{\rm ic}$ and $\sigma \approx -1.5$ if $B > 10B_{\rm ic}$ (see Figure \ref{fig:field_sigma}). The distance measure equation will be stable to variations in the observables for radio source with lobe magnetic field strengths constrained to either of these limiting cases. The distances to sources between these limits can be calculated assuming prior knowledge of the lobe field strength but will have large uncertainties.

The RAiSE dynamical model developed in \citet{Turner+2015} and \citet{Turner+2018a}, although more sophisticated in its treatment of the host environment and lobe dynamics, is expected to have similar dependences to the simpler analytical model used here in this study. However, the analytic model is limited to radio lobes which spend most of their evolutionary history in the inner regions of the cluster \citep[before the second break in the density profile; e.g.][]{Vikhlinin+2006, Shabala+2008} where the gas density can be better approximated by a single power-law. Throughout this work this analytic form will be used for clarity to the reader, however it should be noted that marginally improved results can be obtained by using the full RAiSE model, especially for sources expanding further into the cluster environment.

\section{Absolute distance to Cygnus A}
\label{sec:Distance to Cygnus A}

We test the validity of the AGN standard candle distance measure technique developed in the previous sections by applying this method to the well-studied radio galaxy Cygnus A. This object has the required set of observations for this technique, as outlined in Section \ref{sec:High-redshift radio source catalogue}. The distance to Cygnus A is estimated by constructing standard candles, providing a constraint on the Hubble constant, and thus a test of this method by comparing to independent measurements.

\subsection{Radio source and environment properties}

Cygnus A is a low-redshift radio source \citep[spectroscopic redshift of $z = 0.056075 \pm 0.000067$;][]{Owen+1997} with a double FR-II lobe morphology; these lobes are termed the eastern and western lobes in this work. Published high-resolution multi-frequency radio observations provide precise flux densities, linear sizes and volumes for each lobe \citep[e.g.][]{Carilli+1996, Steenbrugge+2010}. The axis ratios of each lobe is estimated using the linear size and volume of the lobe measured by \citet{Steenbrugge+2010} assuming the lobes are ellipsoidal in shape as for the analytical model. This yields axis ratios of 2.8 and 3.02 for the eastern and western lobes respectivelty.

The multi-frequency flux densities can further be used to fit properties of the electron energy distribution ($s$, $\nu_{\rm b}$ and $\gamma$). \citet{Steenbrugge+2010} has core and hotspot removed measurements of the flux density in the east and west lobes of Cygnus A at six frequencies from $151\rm\, MHz$ to $15\rm\, GHz$. These radio spectra for both lobes are fitted using the continuous injection (CI) model following the method of \citet{Turner+2018b}. The spectra of both lobes are fitted with a single injection index of $s = 2.42\pm0.13$ whilst the break frequency is fitted separately for each lobe; the eastern lobe has a spectral break at $0.96\rm\, GHz$ and the western lobe has a break at $1.3\rm\, GHz$.

X-ray observations of the Cygnus A and its cluster environment provide measurements of the gas density at the end of each lobe and an estimate of the equipartition factor in the lobes. \citet{Wellman+1997} found the magnetic field strength in Cygnus A is a factor of $0.24\pm0.02$ times lower than its minimum energy field strength, which is in turn a factor of 0.96 below the equipartition value. The equipartition factor is thus found to be $q = 0.0043\pm 0.0013$ using the relation $B/B_{\rm eq} = q^{2/(s+5)}$ \citep{Croston+2005}.
The de-distanced gas density $\eta$ (introduced in Section \ref{sec:Measuring the host cluster environment}) is derived from the gas densities derived by \citet{Ito+2008} at the end of the radio lobes using literature X-ray observations from \emph{ROSAT PSPC} \citep{Reynolds+1996} and \emph{Chandra} \citep{Smith+2002}. These gas densities are de-distanced using Equation \ref{kband_density2} to remove their assumed cosmology and by assuming a typical slope for the gas density profile of $\beta = 0.82\pm0.15$ \citep{Vikhlinin+2006}.

The observables required by the distance measure equation are listed in Table \ref{tab:summary} for both the east and west lobes of Cygnus A.

\begin{table*}
\begin{center}
\caption[]{Observed and derived parameters for the active radio galaxy Cygnus A. The first three rows tabulate the directly observed properties of the radio source, the next three are parameters derived from the radio spectrum, and the remaining two rows list X-ray measurements of the cluster gas density. The table includes $1\sigma$ measurement uncertainties where known.}
\label{tab:summary}
\renewcommand{\arraystretch}{1.1}
\setlength{\tabcolsep}{8pt}
\begin{tabular}{ccrlrlp{4.5cm}}
\hline\hline
Parameter&Symbol&\multicolumn{2}{c}{East lobe}&\multicolumn{2}{c}{West lobe}&Reference
\\
\hline
axis ratio	&	$A$&	$\sim$2.80\!\!\!\!\!&&	$\sim$3.02\!\!\!\!\!&&-- derived from \citet{Steenbrugge+2010}.\\
linear size	&	$\theta$&	$58.6$\!\!\!\!\!&\!\!\!\!\!$\rm arcsec$&	$67.3$\!\!\!\!\!&\!\!\!\!\!$\rm arcsec$&-- \citet{Carilli+1996}.\\
flux density ($151\rm\, MHz$)	&$S_{151}$&5960$\pm$450\!\!\!\!\!&\!\!\!\!\!$\rm Jy$&4750$\pm$350\!\!\!\!\!&\!\!\!\!\!$\rm Jy$&-- \citet{Steenbrugge+2010}.
\\
\hline
break frequency	&	$\nu_{\rm b}$&	$0.96$\!\!\!\!\!&\!\!\!\!\!$\rm GHz$&	$1.3$\!\!\!\!\!&\!\!\!\!\!$\rm GHz$&\multirow{2}{4.5cm}{-- derived from \citet{Steenbrugge+2010} using \citet{Turner+2018b}.}\\
injection index	&	\!$s = 2\alpha_{\rm inj} + 1$\!&	\multicolumn{4}{c}{2.42$\pm$0.13}&\\
equipartition factor	&	$q$&\multicolumn{4}{c}{0.0043$\pm$0.0013}&-- derived from \citet{Wellman+1997}.\\
minimum electron energy	&$\gamma$	&288$\pm$13\!\!\!\!\!	&&297$\pm$14\!\!\!\!\!	&&-- derived from \citet{Carilli+1996} and \citet{McKean+2016}.
\\
\hline
density exponent	&	$\beta$&\multicolumn{4}{c}{0.82$\pm$0.15}&-- \citet{Vikhlinin+2006}.\\
gas density (de-dist. at $\theta$)	&	$\eta$&	$2.0\times10^{-11}$\!\!\!\!\!&\!\!\!\!\!$\rm kg\, m^{-2.5}$&	$1.6\times10^{-11}$\!\!\!\!\!&\!\!\!\!\!$\rm kg\, m^{-2.5}$&-- derived from \citet{Ito+2008}.
\\
\hline
\end{tabular}
\end{center}
\end{table*}

\subsubsection{Lobe plasma electron energy distribution}

The electron energy distribution generated by the shock-acceleration of jet particles is expected to lower by approximately an order of magnitude as the plasma expands into the radio lobes. The change in internal energy $U$ of a packet of synchrotron-emitting electrons as it expands adiabatically from the hotspot into the lobe is given by $dU/U = 1 - \mathcal{P}^{(\Gamma - 1)/\Gamma}$, where $\Gamma = 4/3$ is the adiabatic index of the lobe plasma and $\mathcal{P}$ is the ratio of the pressure in the lobe to that in the hotspot. The minimum Lorentz factor is related to the minimum electron energy in the lobe through $U_{\rm min} = \gamma m_{\rm e} c^2$. The minimum Lorentz factor in the lobe can thus be derived from the hotspot value and the lobe-to-hotspot pressure ratio using the following equation:

\begin{equation}
\gamma = \gamma_{\rm hs} \mathcal{P}^{(\Gamma - 1)/\Gamma} ,
\end{equation}

where $\gamma_{\rm hs}$ is the minimum Lorentz factor measured in the hotspot. \citet{Carilli+1996} measure the pressure in the hotspots of Cygnus A as $\sim$$3\times10^{-10}\rm\, Pa$ and the pressure along the radio bridge as $6\times10^{-12}\rm\, Pa$; i.e. a lobe-to-hotspot pressure ratio of $\mathcal{P} \sim 0.02$. The Lorentz factors in the lobe are therefore a factor of $\sim$$0.4$ times less than those at the hotspot of Cygnus A.

\citet{McKean+2016} detected a turnover in the hotspot radio spectra in both lobes of Cygnus A using the \emph{Low Frequency Array} (LOFAR). They fitted a minimum Lorentz factor for hotspot A of $\gamma_{\rm hs} = 791\pm15$ (western lobe) and for hotspot D of $\gamma_{\rm hs} = 766\pm15$ (eastern lobe). These values reduce  to $\gamma = 297\pm14$ and $\gamma = 288\pm13$ respectively as the packets of synchrotron-emitting electrons expand adiabatically into the lower pressure lobes. The Lorentz factors of the electron population are expected to further reduce over time due to the expansion of the radio lobe, however the freshly injected electrons which predominantly contribute to the observed spectrum will not experience significant energy losses \citep{Turner+2018a}.

\subsubsection{Cygnus A lobe magnetic fields}

\citet{Carilli+1996} {estimated} the minimum energy magnetic field strength in the lobes of Cygnus A in both the lobe heads and the radio bridge connecting the hotspots to the core. Their measurements made based on the level of synchrotron radiation generated from the observed lobe volume yield field strengths of $6.5\rm\, nT$ at the lobe heads and $4.5\rm\, nT$ in the radio bridge. \citet{Wellman+1997} found the lobe magnetic field is a factor of $0.24\pm0.02$ less than the minimum energy field strength {(using three dynamical model based techniques)}, leading to estimates of 1.56 and $1.08\rm\, nT$ at the lobe heads and radio bridge respectively. {By contrast, \citet{deVries+2018} used X-ray inverse-Compton measurements to find a field strength of approximately $4\rm\, nT$ at the lobe rim, with lower values of $2.7_{-0.4}^{+0.5}\rm\, nT$ and $1.7_{-0.3}^{+0.7}\rm\, nT$ towards the centre of the eastern and western lobes respectively. The field strengths estimated using both these techniques are broadly consistent, though inconveniently} are within a factor of ten of the field equivalent of the microwave background radiation at this redshift ($B_{\rm ic} = 0.35\rm\, nT$); {the precision of the distance measurements for Cygnus A will thus} largely depend upon the precision of the magnetic field strength {estimates}. This will not be true for much weaker, or slightly more powerful sources, which would be better targets for the AGN standard candle technique.

The magnetic field strength in each lobe is calculated independently in this paper using the same measurements for Cygnus A as catalogued in Table \ref{tab:summary} to ensure a self-consistent solution. {The X-ray inverse-Compton measurements provide the most direct estimate of the magnetic field in Cygnus A, however we refrain from using these field strengths as: (1) some parameters assumed in their calculation are not consistent with the set of observations used in our work; and (2) this technique is not generally applicable since X-ray measurements are not available for most other sources.} The minimum energy magnetic field {\citep[and thus equipartition field to within a few percent;][]{Beck+2005}} is related to the flux density $S_\nu$, lobe volume $V$ and properties of the electron energy distribution through Equation 6 of \citet{Pyrzas+2015}, see also \citet{Longair+1994}:

\begin{equation}
B_{\rm min} = \left[\frac{\mu_0}{V} \frac{1}{\gamma^{s-2}} \frac{m_{\rm e}c^2}{s-2} \frac{L_{\rm \nu}}{g_1(s)\nu^{-(s - 1)/2}} \right]^{2/(s + 5)} ,
\label{Bmin}
\end{equation}

where the $g_1$ is a constant function of the electron energy injection index given by,

\begin{equation}
\begin{split}
g_1(s) &= 2.344 \times 10^{-25} \left[1.253 \times 10^{37} \big(m_{\rm e} c^2 \big)^{2} \right]^{(s - 1)/2} \\
&\quad\times\, \frac{\sqrt{\pi}}{2} \frac{\:\! \Gamma \big(\tfrac{s}{4} + \tfrac{19}{12} \big)\:\! \Gamma \big(\tfrac{s}{4} - \tfrac{1}{12} \big)\:\! \Gamma \big(\tfrac{s}{4} + \tfrac{5}{4} \big)}{(s + 1)\:\! \Gamma \big(\tfrac{s}{4} + \tfrac{7}{4} \big)} .
\end{split}
\label{g1}
\end{equation}

The constants in this equation are in SI units and $\Gamma$ is the Gamma function.

The minimum energy magnetic field strength in the eastern and western lobes of Cygnus A is calculated using Equation \ref{Bmin} assuming the flux densities and lobe volumes of \citet{Steenbrugge+2010}, the electron energy injection index fitted to their flux densities following \citet{Turner+2018b}, and typical minimum Lorentz factors for the lobe as derived in the previous section. This yields minimum energy field strengths of $4.6\pm0.1\rm\, nT$ in the eastern lobe and $4.1\pm0.1\rm\, nT$ in the western lobe, and magnetic field strengths of $1.13\pm0.10\rm\, nT$ and $0.99\pm0.08\rm\, nT$ respectively using the relation found by \citet{Wellman+1997}. {The magnetic field strengths estimated for both lobes are within a factor of two of the X-ray inverse-Compton measurements made by \citet{deVries+2018}, confirming the distance measurements for Cygnus A should be somewhat robust despite its field strength lying close to the microwave background value (deriving $\sigma$ assuming the higher inverse-Compton field strengths leads to an $\sim$$30\%$ decrease in the distance estimates).}

Note that the lobe magnetic field strength estimates have a weak dependence on cosmology (i.e. $\sim$$h^{2/7}$), calculated here assuming the cosmological parameters of \citet{Planck+2016}. These estimates are only used in this work to provide a rough estimate of $\sigma$ and $\kappa$ to demonstrate the precision that could be expected from the absolute distance technique when used in the high or low field strength limits (i.e. $B \gg B_{\rm ic}$ or $B \ll B_{\rm ic}$).

\subsection{Distance measurements}

The distance to Cygnus A is calculated using the transverse comoving distance equation developed in Sections \ref{sec:Constructing radio AGN standard candles} and \ref{sec:Measuring the host cluster environment} for the observations compiled earlier in this section. The magnetic field strengths in the lobes of Cygnus A have been shown to lie within a factor of ten of the field equivalent to the cosmic microwave background radiation potentially leading to instability in the solution. The distance measure is therefore calculated for a range of lobe magnetic field strengths including the high lobe field strength limit ($B \gg B_{\rm ic}$), {estimates} at the lobe heads and radio bridge from the literature \citep{Carilli+1996}, and those for each lobe made in this work using a consistent set of observations. The equipartition factor is the largest source of statistical uncertainty in the distance measurements; this leads to correlated errors for the two lobes of Cygnus A. The distances measured to Cygnus A and the corresponding Hubble parameter fitted assuming a spectroscopic redshift of $z = 0.056075$ are listed in Table \ref{tab:summary2}.

\begin{table*}
\begin{center}
\caption[]{Derived distances to the two lobes of Cygnus A assuming different values for the magnetic field strength. The method is relatively unstable for lobe field strengths close to the equivalent field strength of the cosmic microwave background.}
\label{tab:summary2}
\renewcommand{\arraystretch}{1.1}
\setlength{\tabcolsep}{8pt}
\begin{tabular}{cccccrlrl}
\hline\hline
\multicolumn{4}{c}{Lobe magnetic field properties}&\!\!\!\!\!&\multicolumn{2}{c}{Distance ($d_{\rm M}$)}&\multicolumn{2}{c}{Hubble parameter ($h$)}\\
Location&$B$&$\sigma$&$\kappa$&\!\!\!\!\!&East lobe&West lobe&East lobe&West lobe
\\
\hline
$B \gg B_{\rm ic}$ limit	&	$>$$3.5\rm\, nT$&	-1.5&	1&\!\!\!\!\!&		$174_{-42}^{+55}\rm\,Mpc$&		$141_{-34}^{+44}\rm\,Mpc$&	$0.95_{-0.23}^{+0.30}$&	$1.18_{-0.28}^{+0.37}$
\\
lobe heads	&	$1.56\rm\, nT$&	-1.4&	0.38&\!\!\!\!\!&		$212_{-49}^{+63}\rm\,Mpc$&		$170_{-39}^{+50}\rm\,Mpc$&	$0.78_{-0.18}^{+0.23}$&	$0.98_{-0.22}^{+0.29}$
\\
radio bridge	&	$1.08\rm\, nT$&	-1.3&	0.14&\!\!\!\!\!&		$286_{-62}^{+80}\rm\,Mpc$&		$227_{-50}^{+63}\rm\,Mpc$&	$0.58_{-0.13}^{+0.16}$&	$0.73_{-0.16}^{+0.20}$
\\
\hline
East lobe	&	$1.13\rm\, nT$&		-1.32&	0.17&\!\!\!\!\!&		$271_{-60}^{+77}\rm\,Mpc$&		...&	$0.61_{-0.13}^{+0.17}$&	...
\\
West lobe	&	$0.99\rm\, nT$&	-1.27&	0.10&\!\!\!\!\!&		...&		$251_{-54}^{+69}\rm\,Mpc$&	...&	$0.66_{-0.14}^{+0.18}$
\\
\hline
\end{tabular}
\end{center}
\end{table*}

The distance to the lobes of Cygnus A is {constrained} between 141 and $286\rm\, Mpc$ for the full range of magnetic field strengths considered. The spread in the distance measurements greatly reduces to between 227 and $286\rm\, Mpc$ if only {the estimates} assuming field strengths appropriate for the whole lobe are included (i.e. the radio bridge and averages for the eastern and western lobes). The precision of the technique can be assessed by the self-consistency of the {distances} for the two lobes. The east and west lobes have flux densities, linear sizes and break frequencies differing by 25\%, 15\% and 35\% respectively; consistent distance measurements at a greater precision are only expected if the technique works as intended. The difference in the distances for the eastern and western lobes is $59_{-29}^{+33}\rm\, Mpc$ assuming the magnetic field strength {estimated} at the radio bridge by \citet{Carilli+1996}. The distances to these two lobes are inconsistent with each other, however the lobes {have} different magnetic field strengths \citep{deVries+2018}. Considering the magnetic field strengths estimated for each lobe yields {distances} of $271_{-60}^{+77}\rm\, Mpc$ and $251_{-54}^{+69}\rm\, Mpc$ for the east and west lobes respectively. This $20_{-29}^{+33}\rm\, Mpc$ difference, or equivalently a 7\% fractional error, is far less than the differences seen between the observables for each lobe confirming the method accurately measures the relative distance to radio sources given high-quality data (i.e. different objects at the same distance are correctly measured to have the same distance). 

The ability for AGN standard candles to measure absolute distances can be verified by calculating the Hubble constant using these measurements and comparing to accepted values from other techniques. The Hubble constant is measured using the spectroscopic redshift to Cygnus A ($z = 0.056075$) and assuming typical values for the dark energy and matter densities \citep{Planck+2016}. The four {distances estimated} to the lobes of Cygnus A assuming the field strength at the radio bridge from \citet{Carilli+1996} or the averages for each lobe calculated in this work lie in the range $H_0 = 61$-$73\rm\, km\,s^{-1}\, Mpc^{-1}$. The Hubble constants of $H_0 = 61_{-13}^{+17}$ and $66_{-14}^{+18}\rm\, km\,s^{-1}\, Mpc^{-1}$ are estimated for the eastern and western lobes using the robustly derived magnetic field strengths based on the consistent set of observables. The statistical uncertainty in the Hubble constant, propagated through from the measurement uncertainties (listed in Table \ref{tab:summary2} and discussed in text), mostly results from the moderate uncertainty in the equipartition factor for Cygnus A. The instability of the method due to the closeness of the lobe magnetic field strength to the microwave background equivalent contributes a systematic error of $\pm$$5\rm\, km\,s^{-1}\, Mpc^{-1}$ to the quoted uncertainties. 

These AGN standard candle estimates of the Hubble constant are consistent with other low-redshift techniques including Cepheid variables and type 1a supernovae with measurements by \citet{Riess+2011} of $H_0 = 73.8 \pm 2.4\rm\, km\,s^{-1}\, Mpc^{-1}$. The AGN standard candle point estimates for the Hubble constant are in closer agreement with the measurements made using \emph{Planck} CMB lensing and temperature data which find $H_0 = 67.8 \pm 0.9\rm\, km\,s^{-1}\, Mpc^{-1}$ \citep{Planck+2016}, though the analysis of at least several additional sources would be required to provide support for the value from either technique.

\section{Selecting candidate standard candles}
\label{sec:High-redshift radio source catalogue}

The construction of AGN standard candles using the analytic framework developed in Sections \ref{sec:Constructing radio AGN standard candles} and \ref{sec:Measuring the host cluster environment} requires observations of FR-IIs with at least measurements of: the source size; flux densities across a broad frequency range; the lobe axis ratio; and an approximate measure of the cluster gas density. The other observables, including the equipartition factor and minimum Lorentz factor of the electron energy distribution, generally show minimal variance between sources and can reasonably be modelled using a population average. 

\begin{figure}
\begin{center}
\includegraphics[width=0.47\textwidth,trim={10 7.5 0 10},clip]{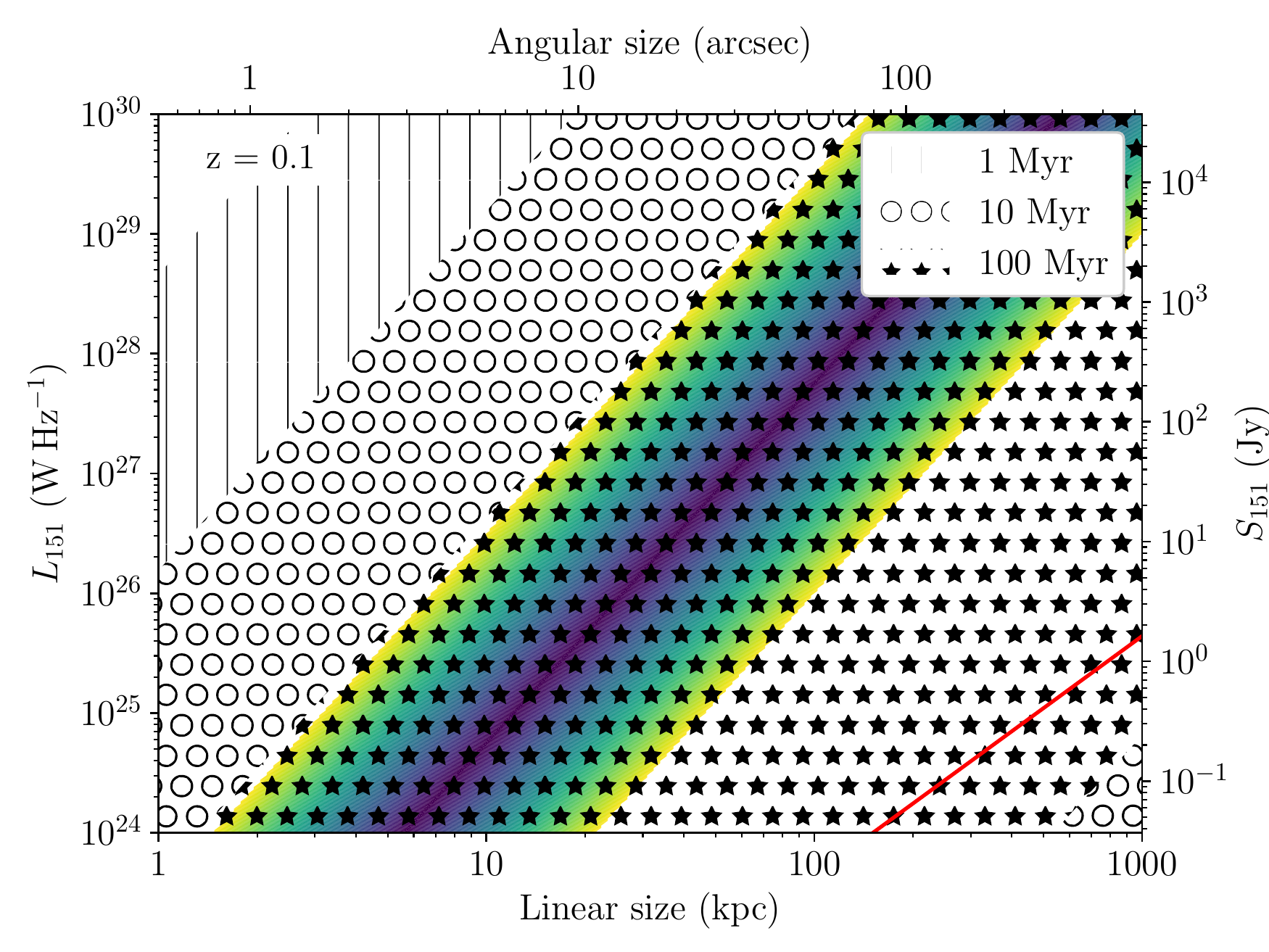} \\
\includegraphics[width=0.47\textwidth,trim={10 7.5 0 7.5},clip]{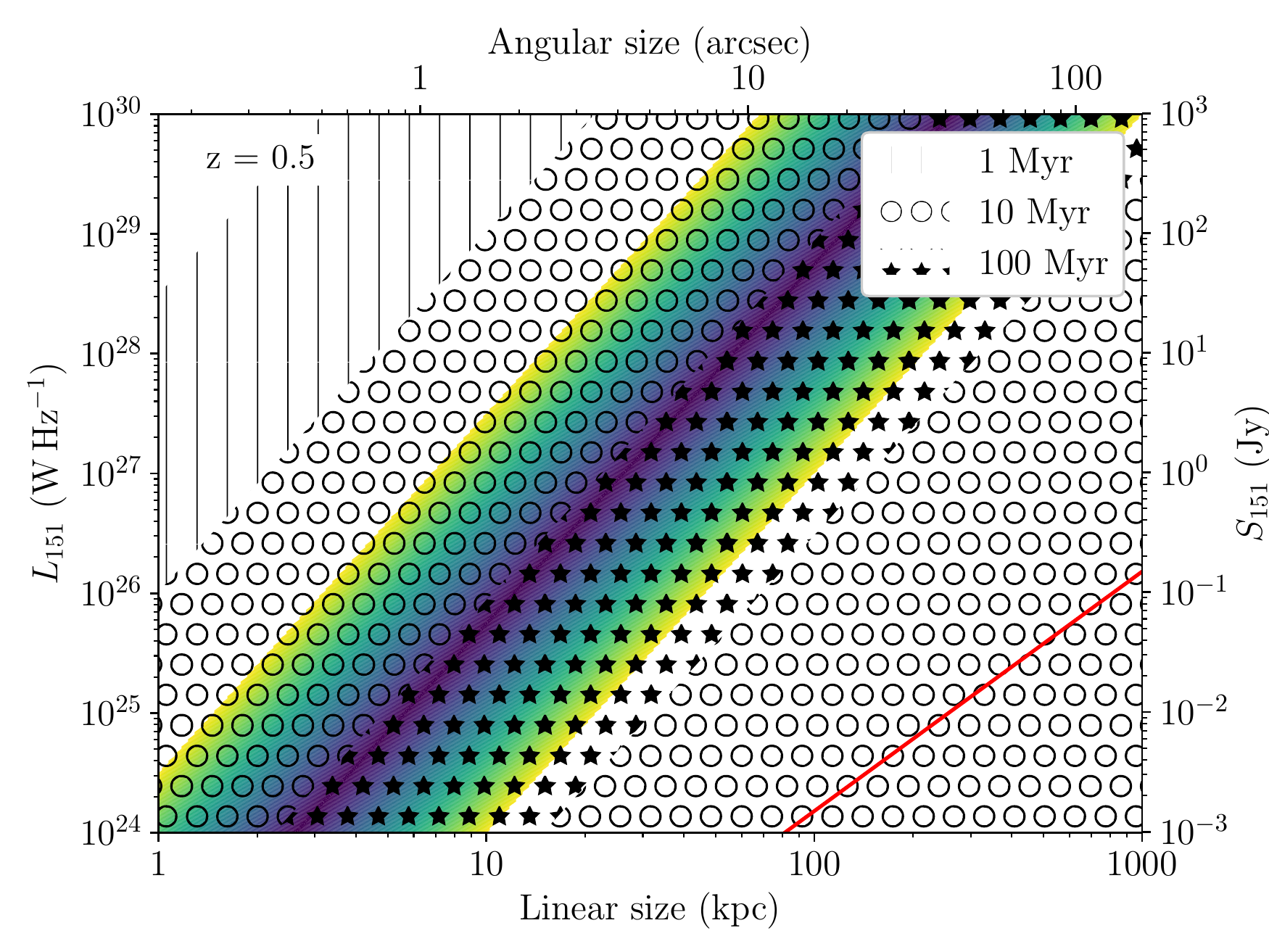} \\\includegraphics[width=0.47\textwidth,trim={10 10 0 7.5},clip]{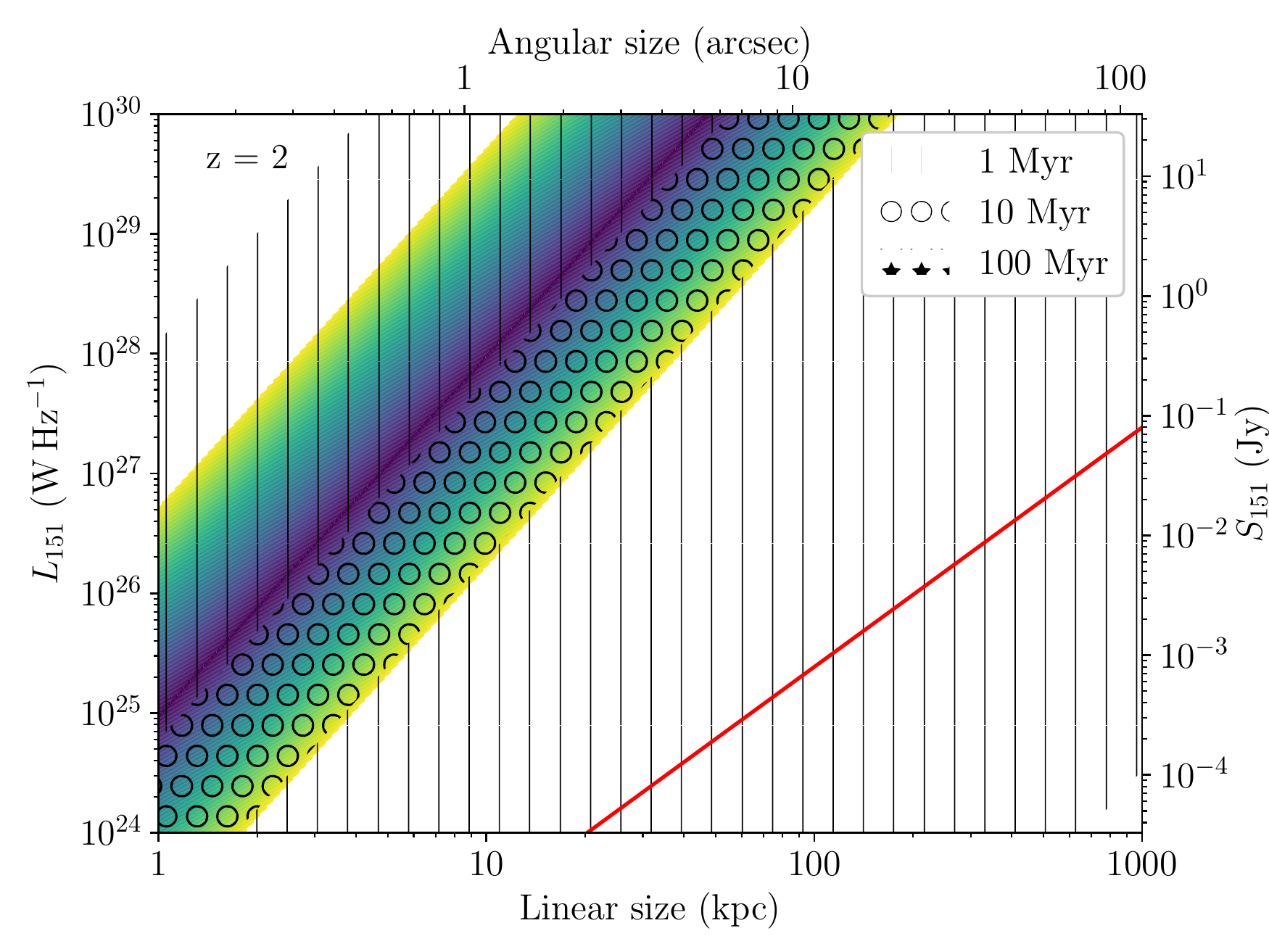}
\end{center}
\caption{Region of flux density--angular size parameter space where radio lobes of age 1, 10 and $100\rm\, Myr$ (hatched lines, circles and stars respectively) have their break frequency in a 0.1 to $10\rm\, GHz$ observing frequency range. The shading plots the fractional difference between the lobe magnetic field strength and the field equivalent to the cosmic microwave background radiation (CMB). The dark violet shading corresponds to equal values in these two magnetic fields (i.e. $B = B_{\rm ic}$). The shading is plotted for lobe field strengths within a factor of three of the CMB value; this region is too unstable for constructing standard candles even with an estimate of the lobe field strength. Higher lobe magnetic field strengths are towards the top-left of the plot and lobe lower field strengths are towards the bottom-right. The red line shows the $5\sigma$ surface brightness detection limit for LoTSS \citep{Shimwell+2019}. The top, middle and bottom panels show the locations in parameter space where radio lobes of each age are detectable at redshifts $z = 0.1$, 0.5 and 2 respectively.}
\label{fig:redshift_dist}
\end{figure}

The brightnesses and sizes of radio lobes with a clearly defined break frequency in the range of typical observing frequencies (e.g. 0.1 to $10\rm\, GHz$) is investigated for a set of mock radio lobes with similar properties to Cygnus A. Specifically, we assume here that the axis ratio, electron energy injection index, minimum Lorentz factor and equipartition factor of Cygnus A are typical for the entire population. Variations in the jet power and density of the ambient medium lead to radio sources taking different evolutionary tracks (i.e. parametric function with time) through the parameter space. The break frequency is estimated with Equation \ref{spectral} assuming a typical source age and by deriving the lobe magnetic field strength as a function of size and luminosity using Equation \ref{Bmin}. The region of flux density--angular size parameter space where such radio lobes with ages of 1, 10 and $100\rm\, Myr$ have a break frequency in a typical 0.1 to $10\rm\, GHz$ frequency range is shown in Figure \ref{fig:redshift_dist} for three redshifts. 

Young-compact and old-extended sources have lobe magnetic field strengths in the strong or weak field limit at low redshifts (i.e. $z \lesssim 0.5$), enabling these radio galaxies to be used to construct standard candles. By constrast, young extended objects (i.e. with high jet powers or underdense environments) make good candidates for standard candles at higher redshifts. This constraint does not greatly cut the high-redshift sample since, in general, only radio sources younger than $10\rm\, Myr$ are detectable at these redshifts due to the strong inverse-Compton upscattering of CMB radiation \citep{BR+1999}.

{Current} surveys such as the \emph{LOFAR Two-metre Sky Survey} (LoTSS) have sufficiently low surface brightness detection limits that only the faintest FR-IIs at high-redshift are excluded from this parameter space \citep[FR-Is more prevalent for $L_{151} < 10^{26}\rm\, W\, Hz^{-1}$;][]{FR+1974}. The LoTSS survey can resolve sources with sizes above 6 arcseconds \citep{Shimwell+2019}, corresponding to a minimum measureable lobe linear size of 3 arcseconds. This constraint only has an effect on the selection of candidate standard candles at moderate redshifts; the other selection criteria have already excluded compact sources at high-redshift.

\section{CONCLUSIONS}
\label{sec:CONCLUSIONS}

We have presented an analytic extension to the RAiSE radio source evolution model \citep{Turner+2015, Turner+2018a} to measure the distances to radio-loud AGNs. This technique uses observations of the integrated flux density, lobe size and axis ratio, and the injection spectral index and break frequency fitted from multi-frequency observations, to construct standard candles enabling the transverse comoving distance to the radio source to be measured. The equation used to construct the AGN standard candles can be modified for several applications including:

\begin{itemize}
\item absolute standard candles at low-redshift ($z \ll 1$) to constrain the Hubble constant;
\item relative distance measurements at high-redshift ($z \gg 1$) to fit the curvature of the universe and determine the dark energy and matter densities;
\item radio continuum redshift measurements of AGNs detected in large-sky radio surveys.
\end{itemize}

The analytical AGN standard candle theory is combined with X-ray observations of the cluster gas density and internal lobe conditions in Cygnus A to construct absolute standard candles. The relative distance measurements will be applied to a sample of high-redshift radio sources in an upcoming publication to constrain the dark energy and matter densities.

The precision of the AGN standard candle technique is demonstrated through the self-consistency of the distance measurements for the two lobes of Cygnus A. The east and west lobes have flux densities, linear sizes and break frequencies differing by 25\%, 15\% and 35\% respectively, yet the distances calculated in this work for these lobes are consistent to within a factor of 7\%. That is, two quite dissimilar radio lobes at the same redshift are correctly identifed to be located at the same distance. These two lobes also have distance measurements consistent with that expected from the host galaxy spectroscopic redshift assuming a $\Lambda$CDM model \citep{Planck+2016}.
The distance estimates for the two lobes together yield a transverse comoving distance to Cygnus A of $261_{-55}^{+70}\rm\, Mpc$ corresponding to a Hubble constant of $H_0 = 64_{-13}^{+17}\rm\, km\, s^{-1}\, Mpc^{-1}$.

The Hubble constant measurement derived using Cygnus A is consistent with the Hubble constant derived from Cepheid and type 1a supernovae (SNe), and cosmic microwave background (CMB) and baryon acoustic oscillations measurements based on the growth of anisotropies in the matter and temperature distribution. The distance measurements made using Cygnus A are somewhat unstable due to the similarity of the lobe magnetic field strength and the field equivalent of the cosmic microwave background radiation. The large statistical uncertainties in the absolute distance measurement and use of a single source prevent the AGN standard candle technique from favouring either the value of the Hubble constant derived using the Cephied and type 1a supernovae measurements or that found from the CMB and baryon acoustic oscillations. The tension between these techniques can be properly investigated using a large sample of radio sources with magnetic field strengths in the strong or weak magnetic field strength limit (i.e. $B \gg B_{\rm ic}$ or $B \ll B_{\rm ic}$) such as many of those in the \emph{Third Cambridge Catalogue of Radio Sources} (3C); the larger sample will reduce the statistical uncertainies whilst choosing objects with a limiting case magnetic field will largely remove any systematic uncertainty.

\subparagraph{}
RJT thanks the University of Tasmania for an Elite Research Scholarship and the CSIRO for a CASS studentship. SSS thanks the Australian Research Council for an Early Career Fellowship, DE130101399, the Australian Government for an Endeavour Research Fellowship, and the University of Hertfordshire for their hospitality. We thank an anonymous referee for their helpful comments that have improved the manuscript.

\end{document}